\documentclass[12pt,a4paper]{article}
\usepackage{amsmath}
\usepackage{amsfonts}
\usepackage{amssymb}
\usepackage{natbib}
\usepackage{algorithm}
\usepackage{algpseudocode}
\usepackage{amsthm}
\usepackage{graphicx}
\usepackage{mathtools}
\DeclarePairedDelimiter{\norm}{\lVert}{\rVert}

\numberwithin{equation}{section}

\begin{document}

\begin{titlepage}
   \vspace*{\stretch{1.0}}
   \begin{center}
      \Large\textbf{Exact Post-selection Inference For Tracking S\&P500}\\
	  \large\textit{Farshad Noravesh}\footnote{Email:inorvesh@ut.ac.ir}	  
	  \large\textit{Hamid Boustanifar} \footnote{Email:hamid.boustanifar@edhec.edu}      
      
   \end{center}
   \vspace*{\stretch{2.0}}
\begin{abstract}
The problem that is solved in this paper is known as index tracking. The method of Lasso is used to reduce the dimensions of S\&P500 index which has many applications in both investment and portfolio management algorithms. The novelty of this paper is that post-selection inference is used to have better modeling and inference for Lasso approach to index tracking. Both confidence intervals and curves indicate that the performance of Lasso type method for dimension reduction of S\&P500 is remarkably high.\\
Keywords: index tracking, lasso, post-selection inference, S\&P500
\end{abstract}

\end{titlepage}

\section{Introduction}
There are two main types of investment strategies, namely active and passive. In passive investment, the goal is to track an index whereas active strategies aim to beat their index by exploiting trading strategies such as market timing. Index replication or index tracking is one of the main methods in passive investing to mimick market index performance. In active investment, the idea is that markets are not efficient and alpha can be achieved by utilizing these inefficiencies in the market by high frequency rebalancing of the portfolio. One of the drawbacks of active strategies is the high transaction costs and the complex dynamics and prediction of these switchings. Index tracking which is a passive investment can be done either static or dynamic. In static approach, the process is so easy and only commitment to an index provides excellent tracking of index. Although static approach is straightforward, but the main issue is that many of these assets are highly illiquid and also transactions costs will be high if all of them should have some allocations. This is the motivation for dynamic approach in the context of passive management that is considered in the present paper. Thus, only a reduced number of assets that make an index are enough to replicate the index and therefore has low transaction costs and the tradeoff between tracking efficieny and transaction costs should be the main priority and it is an implicit biobjective optimization problem by itself. There are many other advantages to dynamic index tracking which makes it flexible and feasible to develop complex statistical arbitrage as is mentioned in \cite{SantAnna2020}.
A natural tracking error is empirical tracking error (ETE) as is mentioned in \cite{Konstantinos2018} which is defined as :
\begin{equation} \label{eq-ETE}
ETE(w)=\frac{1}{T}\norm{Xw-r_{b}}^{2}_{2}
\end{equation}
where w is designed portfolio. The index that is used in the present paper is S\&P500, however any other index could be used such as the Russell 1000,CSI 300 and Nikkei 225. Many methods and methodology are used in the literature to find the best representative index of an original index for index tracking purposes. There are many regression methods are in the literature for index tracking such as composite quantile regression in \cite{Li2020} or using elastic nets in \cite{Wu2014} for tracking CSI 300 Index and SSE 180 Index by selecting around 30 stocks only.
LASSO method is a powerful algorithm in both regression and dimension reduction. \cite{Tibshirani1996} suggests a method called "Least Absolute Shrinkage and Selection Operator" (LASSO), which is an operator that minimizes the residual sum of squares. The LASSO plays an important role for feature selection. The method arranges the model parameters by shrinking (regularization) the regression coefficients and diminishing some of them to zero. The feature selection takes place after the shrinking phase, each non-zero value is chosen for use in the model. First LASSO method is used for dimension reduction of S\&P500, then post selection inference method used to see which predictors are more effective and less uncertain for their effect on the index and they are called principal components in the context of principal component analysis(PCA). There are many methods to solve the lasso problem as described in \cite{Hastie2015}. \cite{Konstantinos2018} uses majorization minimization (MM) but the present paper uses coordinate descent although the choice of the solver is not the issue for the current paper. Once the LASSO algorithm or any other linear regression method is implemented, the next important step is to see which of these coefficients are more certain in the framework of statistical inference. Exact post-selection inference explained in \cite{Jason2016} is one of the relative recent ideas for inference which unlike traditional approach, inference is done after selection by data. The approach that is used in present paper for post-selection inference(POSI) is called Exact POSI and is described in \cite{Jason2016}. There are other methods for POSI such as data splitting[\cite{Cox1975}] which uses half of data for inference and the other half for selection. Although many statisticians use this method, but it reduces the statistical power since half of data is neglected and inferences are only valid for data that is used in selection. \cite{Tian2018} have introduced randomized response method and have shown it has smaller typeII error in comparison with methods like data splitting or data carving but is computationally challenging.   
\section{Methodology}
\begin{figure}[H]
  \includegraphics[width=\linewidth]{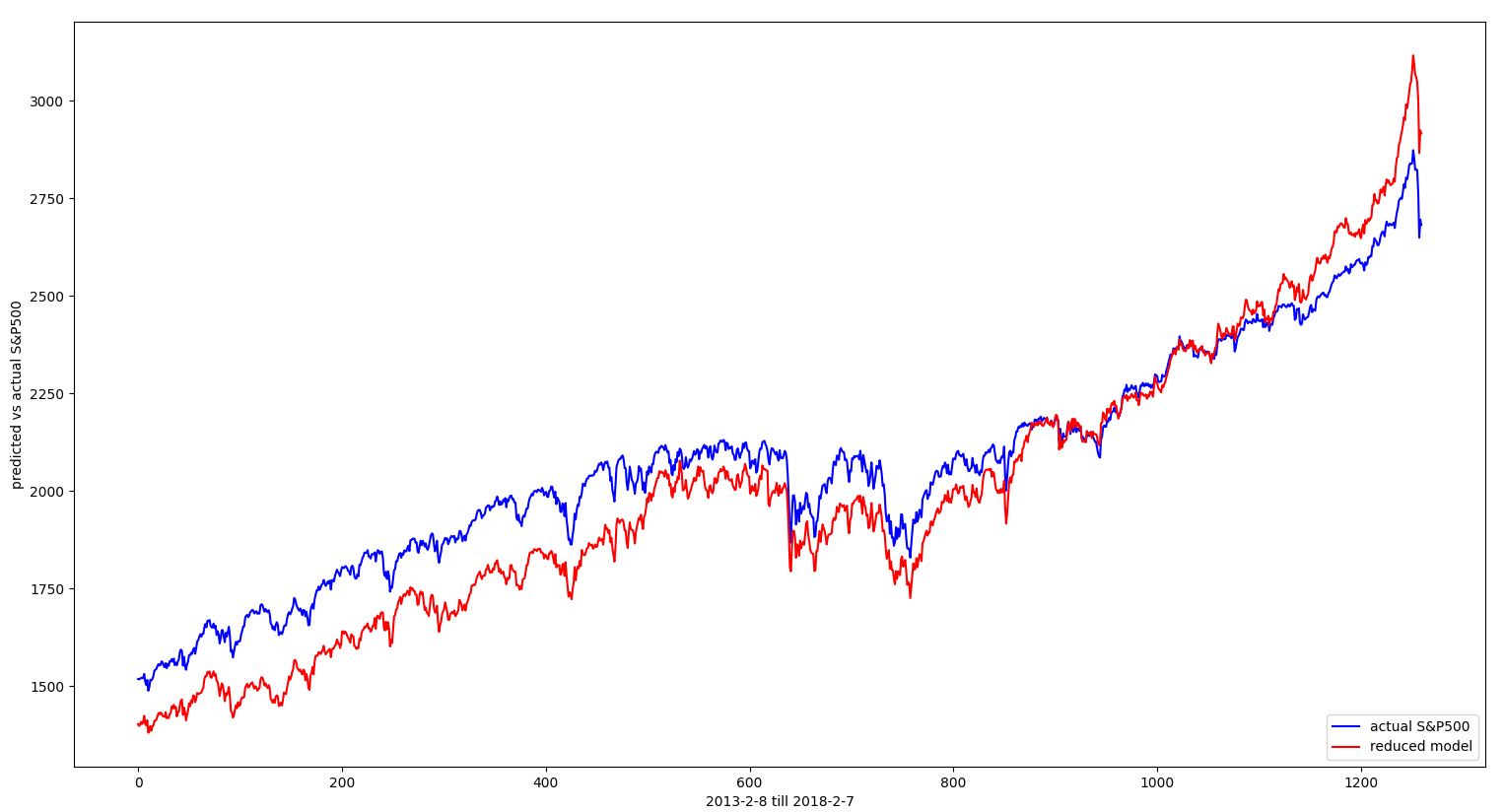}
  \caption{Comparing reduced model and S\&P500 in case 1}
  \label{fig-case1Comp}
\end{figure}

\begin{figure}[H]
  \includegraphics[width=\linewidth]{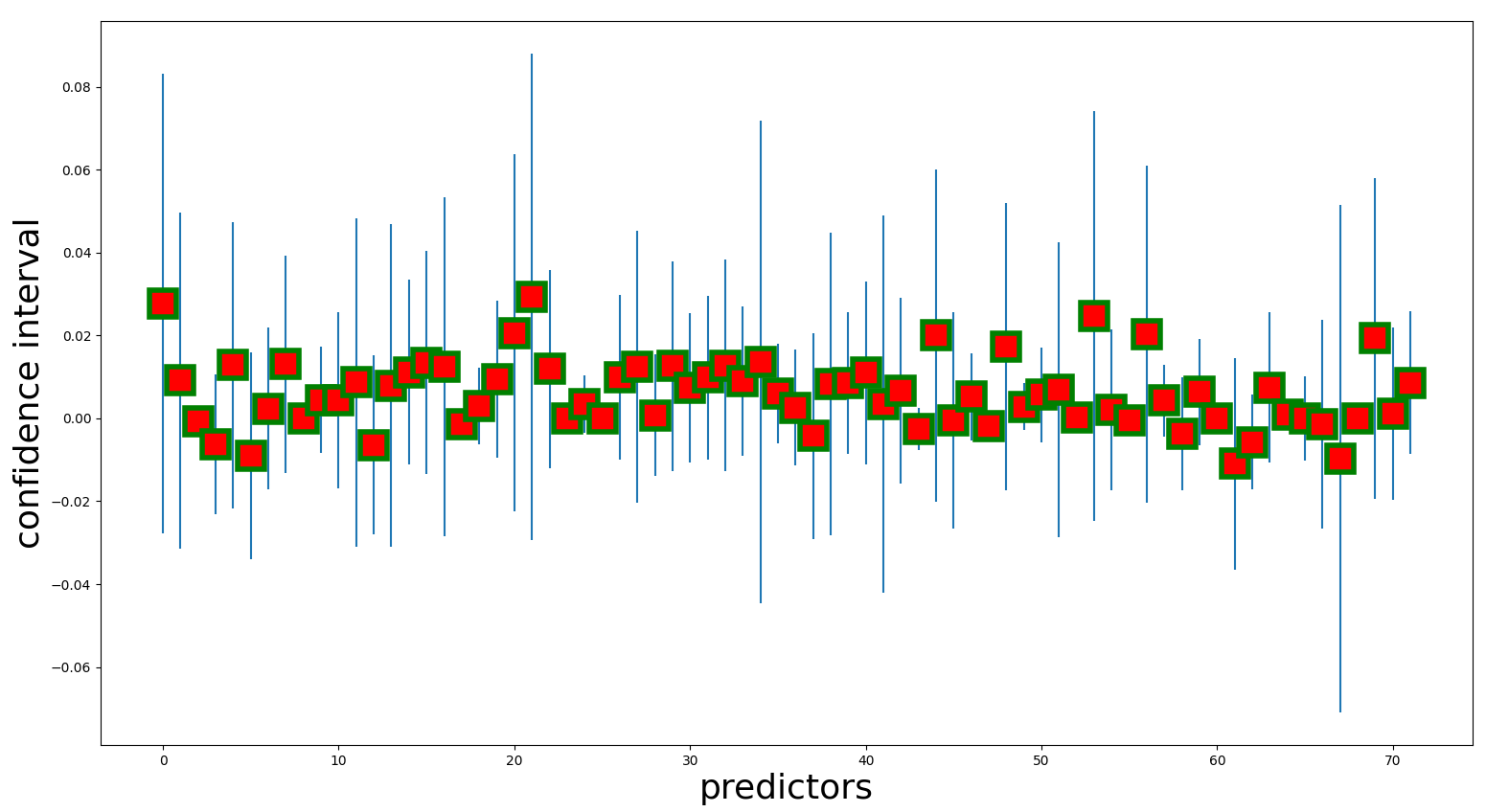}
  \caption{confidence intervals of case 1}
  \label{fig-case1CI}
\end{figure}

\begin{figure}[H]
  \includegraphics[width=\linewidth]{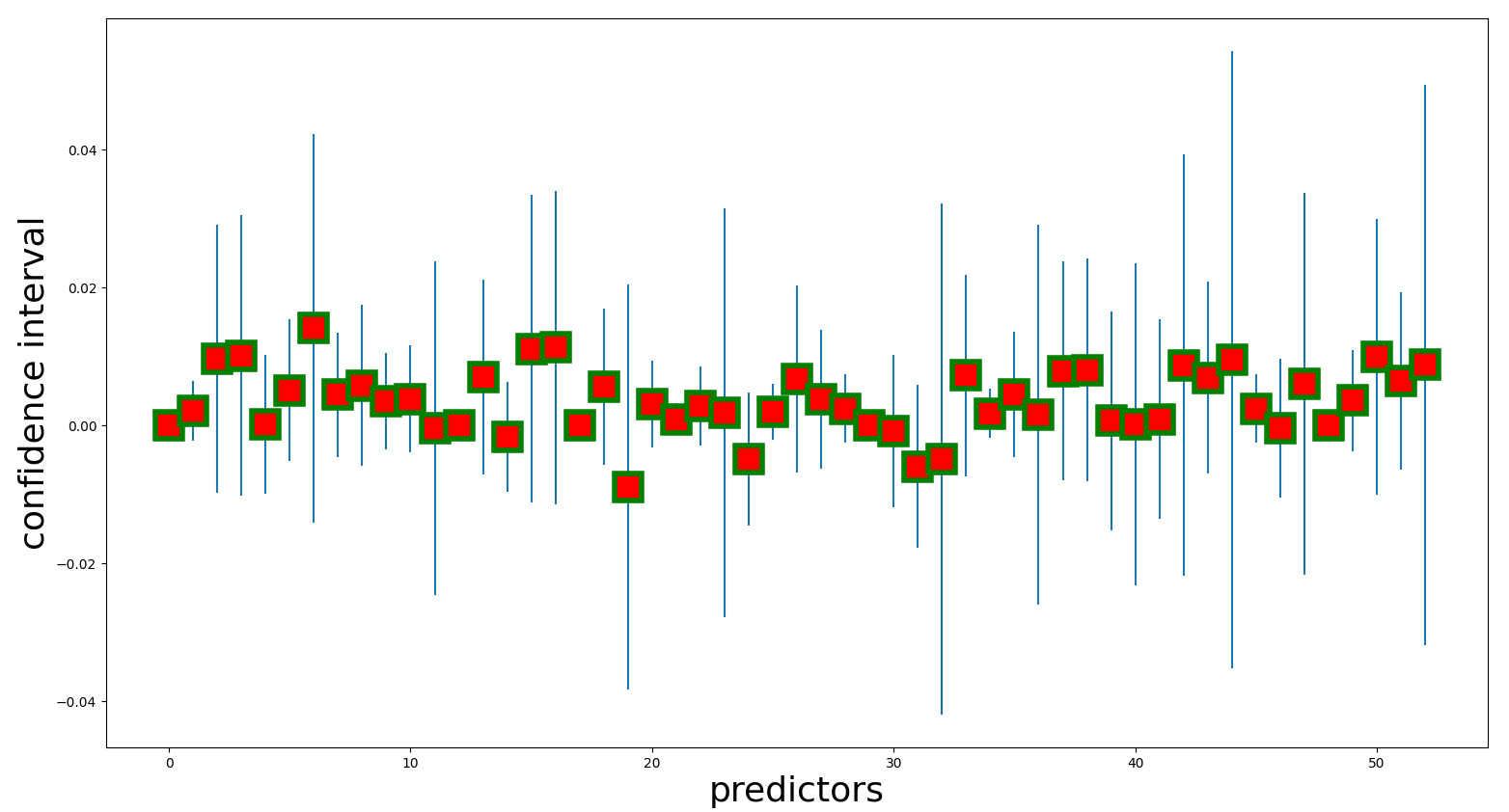}
  \caption{confidence intervals of case 5}
  \label{fig-case5CI}
\end{figure}

The data used in this study is the daily stock prices of 505 equities in the USA Standard S\&P500 index, for the period 02 August till 2013-02 July 2018; which consist of 1259 record. The data is taken from Kaggle. In this paper, the LASSO method [1] is combined with Post-selection inference \cite{Jason2016} to reduce the dimensionality of S\&P500 . \cite{Jason2016},\cite{Hastie2015} shows that the LASSO parameters are bounded by the following interval after selection $Ay\leq b$ is done by
\begin{equation}  \label{eq-polyhedral}
\begin{split}
\{Ay\leq b\}&= \{ \nu^{-}y  \leq  \eta^T y\leq \nu^{+}y  ,\    \nu^{0}y\geq 0 \} \\
\alpha&=\frac{A\eta}{\norm{\eta}^{2}_{2}} \\
\nu^{-}(y)&=max_{ j:\alpha_{j}< 0} \frac{b_{j}-(Ay)_{j}+\alpha_{j}\eta^{T}y}{\alpha_{j}}  \ ,j=1,\hdots p \\
\nu^{+}(y)&=min_{ j:\alpha_{j}> 0} \frac{b_{j}-(Ay)_{j}+\alpha_{j}\eta^{T}y}{\alpha_{j}}  \ ,j=1,\hdots p \\
\nu^{0}(y)&=min_{j:\alpha_{j}=0}(b_{j}-(Ay)_{j}) \ ,j=1,\hdots p
\end{split}
\end{equation}
where y in \ref{eq-polyhedral} is the response variable and is assumed to have a normal distribution in each event as
\begin{equation}
y\approx N(\mu,\Sigma)
\end{equation}
Equation~\ref{eq-polyhedral} is called polyhedral lemma. Ten cases are considered for different values for regularization parameter ($\lambda$) as is shown in table \ref{table-experiments}. Throughout all these cases some parameters are fixed. The first fixed parameter in all these cases is convergence tolerance for Lasso algorithm which is fixed at $0.000001$. The second fixed parameter is maximum iterations in any lasso experiment done in this paper which is $100000$. Finally the third fixed parameter is significant level which is fixed at $0.05$ to bound type-I error in all experiments. Among these 10 cases, only case1 and case5 are compared that correspond to regularization parameter at $0.00002$ and $0.00005$ respectively. n in \ref{table-experiments} is the number of samples and m is the number of datapoints in each event. In order to compare the actual S\&P500 time series with the reduced model, two measures are used. The first one is ETE in \ref{eq-ETE} while the second measure is correlation coefficient between these two time series and is denoted by corr in Tabel \ref{table-experiments}.
 Consider the first case which enabled Lasso, to reduce the number of predictors to only 72 predictors(equities) from 505 equities in S\&P500. The goal of inference is to reduce it even more and therefore only 67 predictors lie in 95 percent confidence interval. By doing this the following test is done for all cases. 
\begin{equation} \label{eq-test}
F^{ [\nu^{-},\nu^{+}] }_{\beta^{M}_{j},\sigma^{2}\norm{\eta}}\approx Unif(0,1)  \ ,j=1,\hdots p
\end{equation} 
which encourages the coefficient's distribution to be a truncated gaussian and remove all coefficients which do not satisfy this test. Thus, the beta coefficients are inside the confidence interval if the CDF of truncated distribution($F$) lies in an interval as is defined in \ref{eq-ci}.
\begin{equation}  \label{eq-ci}
C^{M_j}_{j}=\{\beta^{M}_{j}:\frac{\alpha}{2}\leq F^{ [\nu^{-},\nu^{+}] }_{\beta^{M}_{j},\sigma^{2}\norm{\eta}} \leq 1-\frac{\alpha}{2} \} \ ,j=1,\hdots p
\end{equation} 
\begin{table}[H]
\begin{tabular}{|c|c|c|c|c|c|c|c|} 
\hline
cases & $\lambda$ & n & m & corr & ETE & $p_p$ & $p$ \\ 
\hline\hline
case 1 & 0.000018 & 5 & 30 & 0.992473 & 0.0841 & 67 & 72 \\
\hline
case 2 & 0.000019 & 15 & 10 & 0.984619 & 0.1150 & 67 & 67 \\
\hline
case 3 & 0.00002 & 5 & 10 & 0.980843 & 0.1574 & 54 & 63 \\
\hline
case 4 & 0.00002 & 5 & 5 & 0.980935 & 0.0485 & 59 & 63  \\
\hline
case 5 & 0.000025 & 5 & 5 & 0.975577 & 0.1269 & 48 & 53 \\
\hline
case 6 & 0.00004 & 5 & 10 & 0.965080 & 0.1297 & 26 & 26 \\
\hline
case 7 & 0.000045 & 5 & 5 & 0.923647 & 0.07925 & 22 & 22 \\
\hline
case 8 & 0.000048 & 5 & 5 & 0.954842 & 0.1210 & 16 & 17  \\
\hline
case 9 & 0.00005 & 5 & 10 & 0.872521 & 0.1790 & 16 & 16 \\
\hline
case 10 & 0.00005 & 20 & 5 & 0.971625 & 0.1220 & 16 & 16  \\
\hline 
\end{tabular}
\caption{experiments on using exact POSI for Lasso on S\&P500}
\label{table-experiments}
\end{table}
The process is summarized in Algorithm~\ref{alg:exactPOSI}. To get active set of predictors which are the predictors that have non-zero coefficients, the following LASSO regression is solved:
\begin{equation} \label{eq-LASSO}
min_{\beta\in R^{p}} \{\frac{1}{N}\norm{y-X\beta}^{2}_{2}+\lambda \norm{\beta}_{1}   \}
\end{equation}
$\lambda$ in \eqref{eq-LASSO} is the regularization parameter and is varied in Table~\ref{table-experiments} from $0.000018$ to $0.00005$. As this parameter is increased, the number of selected predictors decreases and therefore less assets are required to track the S\&P500 but this increases the ETE in \ref{eq-ETE}. This can be easily seen that Figure~\ref{fig-case1Comp} has lower error in case 1 and requires 72 assets while case 5 in Figure~\ref{fig-case5Comp} has bigger ETE and requires only 53 assets. Thus,the optimum parameters such as $\lambda$ could be found. The second step in Algorithm~\ref{alg:exactPOSI} finds vector b which is an $n\times 1$ vector and matrix A which is a $p\times n$ matrix. The last step, uses the exact POSI algorithm to find which predictors are more significant with significant level of $0.05$ which is fixed in all these 10 cases. The obtained confidence intervals are shown in Figure~\ref{fig-case1CI} and Figure~\ref{fig-case5CI} for case 1 and case 5 respectively.   
\begin{algorithm}[H]
\caption{exact POSI on Lasso for S\&P500 }
\label{alg:exactPOSI}
Input: given returns for 505 assets and the returns for the S\&P500 index \\
1: run the lasso or ordinary regression or any algorithm to select active set for predictors  \\
2: find b, A such that $Ay\leq b$    \\
3: loop for each predictor:
    for each eta calculate $\nu^{-}$ and $\nu^{+}$ using polyherdal lemma
    calculate $\eta$ and $\beta$ and pick $\beta$ if it satisfies \eqref{eq-ci} otherwise, remove the predictor since it does not have big impact on y
\end{algorithm}

\begin{figure}[H]
  \includegraphics[width=\linewidth]{case5Comp.png}
  \caption{Comparing reduced model and S\&P500 in case 5}
  \label{fig-case5Comp}
\end{figure}

\section{Conclusion}
S\&P500 is one of the targets for passive investors and therefore it is important to discover the main drivers of it. In this paper, the dimension of SPX is reduced using the LASSO method. Exact POSI which is one of selective inference methods is used in the present work to figure out which assets are statistically more significant in driving S\&P500 . Apart from the issues mentioned so far, a bigger picture can be imagined which is inspired by recent interests in index funds and also introducing new indices such as Bloomberg indices. The practitioner who works with the parameters of the algorithm presented in this paper, can introduce and suggest new indices which can mimic the market more efficiently with less number of securities.
\section{Future work}
The present algorithm could be used in higher moment portfolio optimization as well. Since estimating covariance matrix, co-skewness and co-kurtosis for a portfolio of large number of assets is not robust, a good approach is to first discovering which assets are more significant in some responses such as portfolio variance, skewness and kurtosis and then running the multiobjective  portfolio optimization problem over the reduced number of assets.
\bibliographystyle{agsm}
\bibliography{mainDriversofSPX}
\end{document}